\documentclass[aps,prl,twocolumn,groupedaddress]{revtex4} 
\usepackage{epsfig,amsopn}
\begin{document}
%\date{\today}

\title{Non-equilibrium transport through a vertical quantum dot \\
in the absence of spin-flip energy relaxation}

\author{T. Fujisawa$^{1}$}
\altaffiliation[Electronic address:]{fujisawa@will.brl.ntt.co.jp.}
\author{D. G. Austing$^{1}$}
\altaffiliation[Present address:]{National Research Council of Canada. Ottawa, Ontario K1A 0R6, Canada.}
\author{Y. Tokura$^{1}$}
\author{Y. Hirayama$^{1,2}$}
\author{S. Tarucha$^{1,3,4}$}
\affiliation{$^{1}$NTT Basic Research Laboratories, NTT Corporation, 3-1 Morinosato-Wakamiya, Atsugi, 243-0198, Japan}
\affiliation{$^{2}$CREST, 4-1-8 Honmachi, Kawaguchi, 331-0012, Japan}
\affiliation{$^{3}$University of Tokyo, Bunkyo-ku, Tokyo, 113-0033, Japan}
\affiliation{$^{4}$ERATO Mesoscopic Correlation Project, 3-1, Morinosato-Wakamiya, Atsugi, 243-0198, Japan}

%\date{}


\begin{abstract}
We investigate non-equilibrium transport in the absence of spin-flip energy
relaxation in a few-electron quantum dot artificial atom. Novel
non-equilibrium tunneling processes involving high-spin states which cannot
be excited from the ground state because of spin-blockade, and other
processes involving more than two charge states are observed. These
processes cannot be explained by orthodox Coulomb blockade theory. The
absence of effective spin relaxation induces considerable fluctuation of the
spin, charge, and total energy of the quantum dot. Although these features
are revealed clearly by pulse excitation measurements, they are also
observed in conventional dc current characteristics of quantum dots.
\end{abstract}

\maketitle

\vspace*{-3mm}

A quantum dot (QD) is a small conducting island in which electrons occupy
discrete energy states \cite{LPKprogress}. Energy relaxation from
an excited state (ES) to a ground state (GS) inside a QD is significantly
suppressed if a spin flip is required \cite{Khaetskii,Paillard,FujisawaPRB}.
Our measurements on QDs in the Coulomb blockade (CB) regime indicate
an extremely long spin-flip energy relaxation time, $\tau _{spin}>$ 1 $\mu $%
s \cite{FujisawaPRB,FujisawaLV}, which is much longer than the
interval of tunneling events ($\Gamma ^{-1}=$ 1 - 100 ns for a typical
tunneling current of 1 - 100 pA, where $\Gamma $ is the tunneling rate) as
well as the momentum relaxation time, $\tau _{mo}\sim $ 1 ns \cite%
{Bockelmann,JWeis,Agam}. In the absence of efficient energy relaxation,
excess energy remains in the system. This opens up novel non-equilibrium
transport channels, which we describe in this Letter.

We start from the orthodox theory which describes CB and single electron
tunneling (SET). The total energy, $U(N)$, of the system, in which an island
containing $N$ electrons is affected by a gate voltage, $V_{g}$, via a
capacitance, $C_{g}$, is 
\begin{equation}
U(N)=\frac{(-Ne+C_{g}V_{g}+q_{0})^{2}}{2C_{\Sigma }}+E_{int}(N)  \label{EqU}
\end{equation}%
\cite{LPKprogress,SET}. The first term is the electrostatic energy
approximated by a constant Coulomb interaction. Inside the parenthesis is
the sum of the charge on the dot, the induced charge by the gate, and an
offset charge, $q_{0}$. $C_{\Sigma }$ is the total capacitance of the dot.
The second term, $E_{int}(N)$, is the sum of the energies of the occupied $N$
electron levels, measured relative to the Fermi energy of the leads,
accounting for the internal degrees of freedom of the QD. Other corrections
to many-body interactions are also included in $E_{int}$. In the orthodox
theory, originally considered for a continuum density of dot states, the
second term is neglected because {\it the QD is assumed to relax quickly to
the minimum energy, }$E_{int}^{(\min )}${\it , which is independent of }$N$%
{\it .} In this SET scheme, an electron that has entered the island leaves
before another electron is allowed to enter when $U(N_{0})=U(N_{0}+1)$ as
shown in Fig. 1(a). This situation is maintained unless the excitation
energy exceeds the charging energy, $E_{c}\equiv e^{2}/C_{\Sigma }$. In this
picture spin is neglected and $N$-electron GSs are only considered.

\begin{figure}[h]
\epsfxsize=3.5in 
\epsfbox{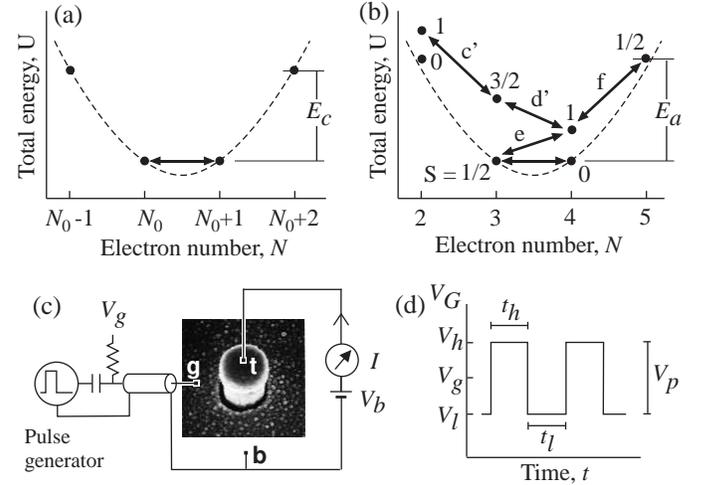}
%\vspace*{.05in}
\caption{(a) Total energy $U(N)$ of a small classical island when SET
allows $N$ to fluctuate between $N_{0}$\ and $N_{0}+1$. (b) $U(N,S)$ of QD-1
at $B=$ 3 T and $V_{G}\sim $ -1.7 V. The arrows indicate allowed tunneling
transitions with excitation energy $<E_{a}$ (see text). (c) Schematic setup
for the pulse measurement on a vertical QD. The circular pillar is of
diameter 0.54 $\protect\mu $m\ (0.50 $\protect\mu $m) for QD-1 (QD-2). See
Ref. \protect\cite{TaruchaAA} for details. All the measurements are
performed at a temperature $\lesssim $ 0.1 K. The magnetic field is applied
parallel to the current. (d) The time-dependent gate voltage $V_{G}(t)$.}
\end{figure}

For a QD in which energy quantization and many-body interactions are
significant, we must consider the discrete energy of the dot, $E_{int}(N,S,M)
$, which is characterized by total spin, $S$, and total angular momentum, $M$
\cite{LPKprogress}. We focus on the regime $\tau _{spin}\gg \Gamma ^{-1}\gg $
$\tau _{mo}$, where spin-flip energy relaxation is effectively absent. This
is the typical condition for a dot weakly coupled to the leads -- the
coupling to the leads is still strong enough to give a measurable current ($%
e\Gamma >$ 1 fA), but weak enough to prevent cotunneling current. {\it If
the QD is excited to any }$N${\it -electron state with a different total
spin from that of the }$N${\it -electron} {\it GS, the ES cannot always
relax to the GS before the QD undergoes a tunneling transition to another }$%
N\pm 1$ {\it electron state.} Successive tunneling transitions force the QD
into highly nonequilibrium configurations. Nonetheless, there is always a
selection rule for the tunneling transition to be satisfied: each tunneling
transition changes $N$ by one and $S$ by one half. Tunneling transitions
which change $S$ by more than one half should be blocked (spin blockade) %
\cite{Weinmann,Ciorga}. Figure 1(b) shows a particular $U(N,S)$ diagram,
which can actually be realized in our QD (see below). Long lived ESs are now
included, and the different spin states have different energies because of
direct Coulomb and exchange interactions\cite%
{TaruchaAA,LPKexcitationAA,TaruchaPRL2000}. The allowed tunneling
transitions indicated by the arrows require an excitation energy smaller
than the addition energy, $E_{a}$. All these transitions can cause the dot
state ($N$, $S$, and $U$) to fluctuate dramatically.

In order to investigate highly non-equilibrium transport, we employ a pulse
excitation technique \cite{FujisawaPRB}, {\it which generates only transient
current associated with long-lived spin states}, on two vertical QDs (QD-1
and QD-2) \cite{TaruchaAA,LPKexcitationAA,TaruchaPRL2000}. Electrons are
confined laterally by an approximate two-dimensional harmonic potential
(confinement energy $\hbar \omega _{0}\sim $ 4 meV for QD-1 and $\sim $ 2.5
meV for QD-2). These samples show qualitatively the same behavior. The $N$%
-dependent addition energy, $E_{a}$ = 2 - 5 meV, of QD-1 clearly reveals a
shell structure \cite{TaruchaAA}. $N$, $S$ and $M$\ can be identified from
the magnetic field, $B$, dependence of the SET current spectrum. Zeeman
splitting is not resolved so we neglected it. A square pulse of amplitude $%
V_{p}$, combined with the static gate voltage, $V_{g}$, is applied to the
gate electrode (g) shown in Fig. 1 (c) \cite{FujisawaPRB}. The time
dependent gate voltage, $V_{G}(t)$, illustrated in Fig. 1(d), is $%
V_{G}=V_{h}\equiv V_{g}+\frac{1}{2}V_{p}$ during the high phase of the
pulse, and $V_{G}=V_{l}\equiv V_{g}-\frac{1}{2}V_{p}$ during the low phase.
A small dc bias voltage $V_{b}=$ 0.15 mV is applied so that electrons are
injected from the bottom contact b ($\Gamma _{b}^{-1}\sim $ 10 ns), and
escape to the top contact t ($\Gamma _{t}^{-1}\sim $ 100 ns). The averaged
dc current, $I$, is measured during the $V_{G}$-pulse irradiation.

Figures 2(a) and (b) show pulse excited current spectra of QD-1 for $N=$ 0 -
5 at $B=$ 3 T. A current peak initially observed in the absence of the $%
V_{G} $-pulse is always split into two peaks of equal height, when the $%
V_{G} $-pulse is applied. Weaker additional peaks indicated by the arrows
are due to transient current through ESs \cite{FujisawaPRB}.\ The pulse
length ($t_{h}=t_{l}=$ 300 ns) is chosen to be much longer than $\tau _{mo}$%
, but much shorter than $\tau _{spin}$. In this case, transient current
through an $N$-electron ES is only observed when the ES has a total spin
different from that of any other lower lying $N$-electron state \cite%
{FujisawaPRB,FujisawaLV}.

\begin{figure}[h]
\epsfxsize=3.5in 
\epsfbox{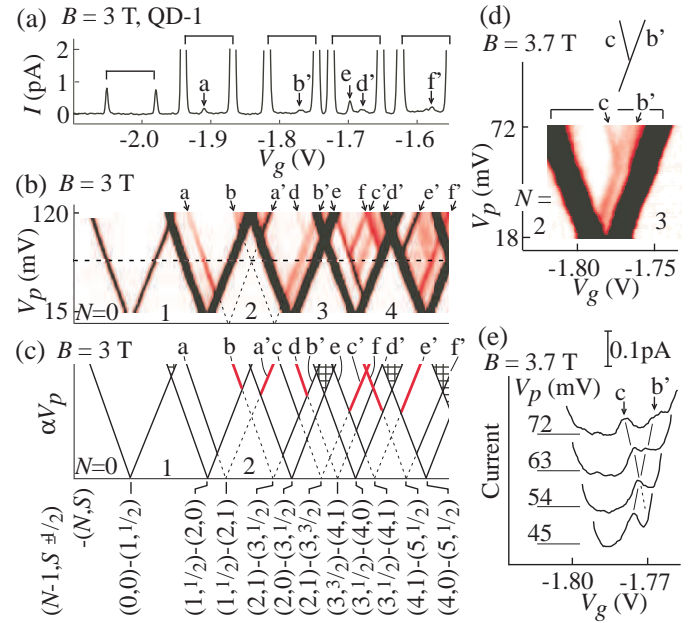}
%\vspace*{.01in}
\caption{ (color). (a) Pulse excited current $I$ vs. gate voltage, $V_{g}$,
for QD-1 measured at $B$ = 3 T and $V_{p}$ $\sim $ 80 mV [dashed line in
(b)]. Each horizontal bar indicates the $V_{g}$ region where $N$ can
fluctuate. (b) Pulse excited current plot at $B$ = 3 T built up from $I$ vs. 
$V_{g}$. The color represents current amplitude [white (0 pA) - red (0.25
pA) - black ($\geq $ 0.5 pA)]. $N$ is fixed in each triangular region
along the bottom due to CB. (c) Representation of current peaks (solid
lines) observed in (b). The transient current peaks are marked by a, a$%
^{\prime }$, b, b$^{\prime }$, etc. The prime (non-prime) indicates
transport during the low (high) phase of the pulse. These peaks are
extrapolated to $V_{p}=0$ V as shown by dashed lines. The tunneling
transitions between $(N-1,S\pm \frac{1}{2})$ and $(N,S)$ states are also
given. The red lines indicate current peaks attributed to the novel DET
process, while cross-hatched regions indicate where normal DET occurs. (d)
Pulse excited current plot [white (0 pA) - red (0.15 pA) - black ($\geq 
$ 0.3 pA)] between $N$ = 2 and 3 at $B$ = 3.7 T. The inset schematically
shows the termination of line c by line b$^{\prime }$. (e) The $I$ vs. $%
V_{g} $ data at various $V_{p}$ values. The curves are offset for clarity. }
\end{figure}

Examples of transient current excited by the $V_{G}$-pulse for $N=$ 1, 2,
and 3, are depicted in Fig. 3(a) (c) and (e), where $U(N,S)$ is plotted
against $V_{G}$. The normal GS-GS tunneling current peaks occur at level
crossings labeled $\circ $, whilst transient current appearing at higher
energy level crossings are denoted by different symbols. Figure 3(a) shows $%
U $ for $N=$ 1 and 2, where a GS-ES tunneling transition labeled $\bullet $
takes place. Consider that a $V_{G}$-pulse is applied as indicated. Suppose
the system is in the GS $(N,S)=(1,\frac{1}{2})$ during the low-phase of the
pulse ($V_{G}=V_{l}$). When the high-phase of the pulse is applied ($%
V_{G}=V_{h}$), the system changes non-adiabatically from $(1,\frac{1}{2})$
at $V_{G}=V_{l}$ to $(1,\frac{1}{2})$ at $V_{G}=V_{h}$ (indicated by arrow
i), since the rise time of the $V_{G}$-pulse (1 - 2 ns) is shorter than $%
\Gamma _{tot}^{-1}$ ($\Gamma _{tot}=\Gamma _{t}+\Gamma _{b}$). If $V_{h}$ is
tuned so that $U(1,\frac{1}{2})=$ $U(2,1)$, transient current can flow at
the crossing marked $\bullet $. Only a few electrons can tunnel through the
ES$(2,1)$, before the inelastic tunneling transition changes the state from $%
(1,\frac{1}{2})$ to $(2,0)$ \cite{FujisawaPRB}. This transient current
appears if the pulse excitation energy, $\alpha V_{p}$, exceeds the level
spacing $\Delta _{2}\equiv U(2,1)-U(2,0)$ as indicated, where $\alpha \equiv
d[U(N)-U(N-1)]/dV_{G}$ is the conversion factor from gate voltage to energy.
The condition for the transient current to flow is given by the thick line
marked $\bullet $ in the $\alpha V_{p}$ - $V_{g}$ plane of Fig. 3(b). This
current is clearly seen as peak a in Fig. 2(b) \cite{higherVp}.

\begin{figure}[h]
\epsfxsize=3.5in 
\epsfbox{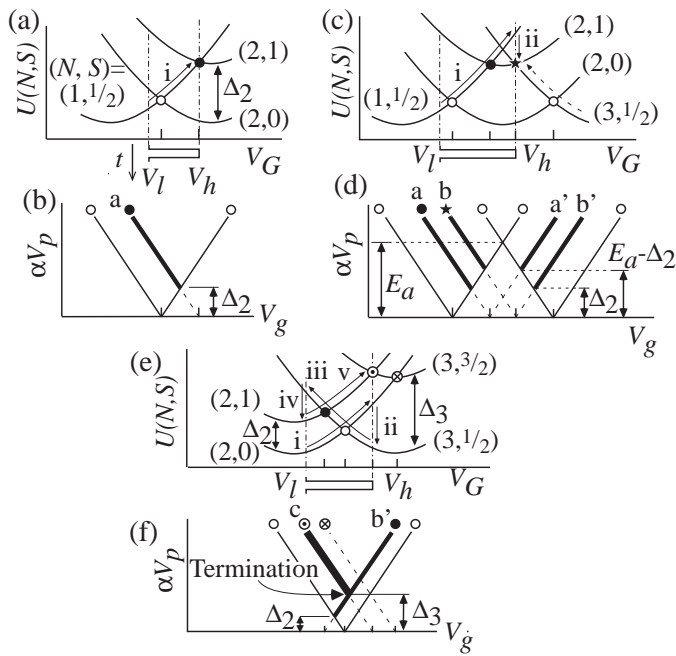}
%\vspace*{.05in}
\caption{ (a), (c), and (e): The gate voltage, $V_{G}(t)$, dependence of the
total energy. The electron number $N$ and total spin $S$ are indicated by $%
(N,S)$. The arrows indicate the excitation/relaxation processes discussed in
the text. The symbols indicate transport processes ($\circ $ for GS-GS
tunneling, $\bullet $ for GS-ES tunneling, $\otimes $ for spin-blockade, $%
\odot $ for ES-ES tunneling, and $\star $ for novel DET). (b),
(d), and (f): Corresponding conditions for stable GS-GS tunneling current
(thin lines) and transient current (thick lines). The horizontal axis is the
static gate voltage, $V_{g}$, while the vertical axis is the excitation
energy of the $V_{G}$-pulse ($=\protect\alpha V_{p}$). }
\end{figure}

Next, we consider striking features for which the conventional SET scheme
collapses. Consider the $U(N,S)$ for the $N=$ 1, 2, and 3 states in Fig.
3(c). Suppose a $V_{G}$-pulse, $\alpha V_{p}<$ $E_{a}$, is applied as shown.
If the state is initially $(1,\frac{1}{2})$ at $V_{G}=V_{l}$, $U(1,\frac{1}{2%
})$ is raised to a higher energy at $V_{G}=V_{h}$ (arrow i), and subsequent
inelastic tunneling results in either the $(2,0)$ or $(2,1)$ state. If it is 
$(2,1)$ (arrow ii), it is now possible for another electron to be
transported at the crossing $U(2,1)=U(3,\frac{1}{2})$ marked by $\star $.
In this case transient transport persists until the inelastic tunneling
transition from $(3,\frac{1}{2})$ to $(2,0)$ occurs, because spin-flip
energy relaxation from $(2,1)$ to $(2,0)$ is absent. The QD returns back to
the initial state $(1,\frac{1}{2})$ when the pulse is switched to $V_{l}$.
This new tunneling process involves three charge states, and thus we
consider it to be novel double electron tunneling (DET). Conventional DET
within the orthodox CB theory appears only for $\alpha V_{p}\geq E_{a}$ %
\cite{LPKprogress,SET}, while the novel DET takes place even for $\alpha
V_{p}<E_{a}$. This process should appear on the thick line b in the $\alpha
V_{p}$ - $V_{g}$ plane of Fig. 3(d).

The novel DET is clearly seen in the experimental data of Fig. 2(b) as\
marked by b. Note that this peak cannot be due to regular SET involving
higher lying $N=$ 2 ESs. The spin singlet ($N=$ 2 GS at $B=$ 3 T) and the
spin triplet [peak a in Fig. 2(b)] are the only possible spin
configurations. So no other ESs with different total spin can appear in the
pulse measurement. Therefore, the pulse measurement successfully allows us
to identify the extra peak b as a novel DET feature. Note that we can also
confirm that peak b is due to novel DET from the $B$-dependence of the peak
position \cite{Bdependence}.

Transient current can also appear during the low phase of the pulse. If a
pulse is applied to excite the QD\ along $(3,\frac{1}{2})$ in the direction
of decreasing $V_{G}$ [dashed arrow in Fig. 3(c)], GS-ES tunneling between $%
(3,\frac{1}{2})$ and $(2,1)$ should appear along line b$^{\prime }$ in Fig.
3(d). Similarly, novel DET appears at the crossing $U(2,1)=U(1,\frac{1}{2})$
[corresponding to feature a$^{\prime }$ in Fig. 3(d)]. Note that the
extrapolated lines a and a$^{\prime }$ (b and b$^{\prime }$) meet at zero
excitation energy. These features are clearly seen in Fig. 2(b).

We now discuss spin-blockade and associated non-equilibrium transport.
Figure 3(e) is the energy diagram for $N=$ 2 and 3 -- the first case where
spin-blockade appears. The direct transition between the GS $(2,0)$ and ES $%
(3,\frac{3}{2})$ is spin-blockaded (marked $\otimes $). However,
non-equilibrium tunneling between the ES $(2,1)$ and ES $(3,\frac{3}{2})$
(marked $\odot $) is allowed if a pulse is applied as indicated. Even
though the QD is initially in the GS $(2,0)$ at $V_{G}=V_{l}$, it first
changes to $(3,\frac{1}{2})$ at $V_{G}=V_{h}$ (arrows i and ii), then to
either $(2,1)$ or $(2,0)$ during the next low-phase [arrows iii and iv for $%
(2,1)$]. If it is $(2,1)$, ES-ES tunneling occurs at $U(3,\frac{3}{2}%
)=U(2,1) $ when the high-phase is restored (arrow v). This kind of tunneling
mechanism leads to the fluctuation of the total spin from 0 to $\frac{3}{2}$%
. Note again that this non-equilibrium fluctuation is attributed to the
absence of effective spin relaxation. This ES-ES tunneling process requires
complex excitation. The excitation energy must be greater than the
corresponding two level spacings: $\alpha V_{p}>\Delta _{2}\equiv
U(2,1)-U(2,0)$ for the excitation indicated by arrow iii, and $\alpha
V_{p}>\Delta _{3}\equiv U(3,\frac{3}{2})-U(3,\frac{1}{2})$ for the
excitation indicated by arrow v. The condition for ES-ES tunneling is marked
by line c ($\odot $) in Fig. 3(f). This line is terminated by line b$%
^{\prime }$ ($\bullet $) for GS-ES tunneling at $V_{G}=V_{l}$, so line c
does not reach the line $\circ $ for GS-GS tunneling. This termination is
the signature of an ES-ES tunneling process which does not involve any GS.

We can identify these features in our QD. Peak c in Fig. 2(d) and (e) now
measured at $B=$ 3.7 T is attributed to ES$(2,1)$ - ES$(3,\frac{3}{2})$
tunneling from the $B$-field dependence of the peak position. This peak is
clearly terminated by peak b$^{\prime }$, which is similarly assigned to ES$%
(2,1)$ - GS$(3,\frac{1}{2})$ tunneling. However, no measurable current ($<$
10 fA) is seen for spin-blockaded tunneling between $(2,0)$ and $(3,\frac{3}{%
2})$ in the region $V_{p}=$ 50 - 100 mV and $B=$ 3.5 - 4.1 T (not shown)
where spin blockade is expected. These observation are consistent with the
above explanation. Note peak c is too weak to see at $B=$ 3 T [Fig. 2(b)].
Nonetheless, the following are very clear in Fig. 2(b) for $N=$ 3 and 4: ES $%
(3,\frac{3}{2})$ - ES $(4,1)$ tunneling line d$^{\prime }$ is terminated by
GS $(3,\frac{1}{2})$ - ES $(4,1)$ tunneling line e, and the termination is
associated with spin-blockaded ES$(3,\frac{3}{2})$ - GS$(4,0)$ tunneling (no
signal).

Other complicated tunneling processes involving both novel DET and ES-ES
tunneling are observed. Peak d (c$^{\prime }$) in Fig. 2(b), which is
terminated by peak b$^{\prime }$ (e), is assigned to ES-ES tunneling, whose
excitation process involves three charge states. The observed pairs of
tunneling lines (c - c$^{\prime }$, d - d$^{\prime }$, etc.) always coincide
when extrapolated to $V_{p}\sim $ 0 V. More complicated excitations are
expected for many-electron QDs. Four different peaks [e, f, c$^{\prime }$
and d$^{\prime }$ in Fig. 2(b)] are observed between the $N=$ 3 and 4 CB\
regions. The corresponding transitions are indicated in the total energy
diagram in Fig. 1(b). The fluctuation in the total energy can be
significantly greater than $\alpha V_{p}$. Non-equilibrium transport can
lead to the accumulation of energy in excess of the excitation energy. As $N$
increases, many\ long-lived ESs with different $S$ can contribute to the
transport. The complexity of many-body excitations increases with $\alpha
V_{p}$ and $N$ \cite{Altshuler}.

\begin{figure}[h]
\epsfxsize=3.5in 
\epsfbox{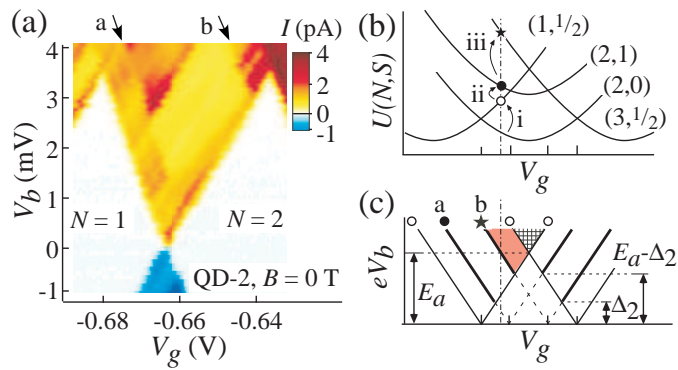}
%\vspace*{.05in}
\caption{ (color). (a) Color plot of the dc current $I$ in the $V_{b}$ - $%
V_{g}$ plane of QD-2 at $B$ = 0 T. The current increases stepwise along the
line marked by arrows a and b, which are assigned to normal GS-ES tunneling
and the novel DET, respectively. (b) Total energy $U(N,S)$ for the dc
excitation processes. (c) Corresponding conditions for the dc excitation
transport in the $eV_{b}$ - $V_{g}$ plane. Novel DET appears in the red
region. }
\end{figure}

{\it Although our findings till now are deduced from pulse measurements, we
expect to see related features in conventional dc excitation measurements.}
For example [see Fig. 4(b)], novel DET involving $N=$ 1, 2, and 3 is allowed
in dc measurements when $eV_{b}$ is greater than $|U(1,\frac{1}{2})-U(2,0)|$
(energy required for normal SET between $N=$ 1 and 2 GSs, arrow i), $%
|U(2,1)-U(1,\frac{1}{2})|$ (excitation energy to the $N=2$ spin-triplet ES
from the $N=1$ GS, arrow ii), and $|U(3,\frac{1}{2})-U(2,1)|$ (extra energy
for novel DET, arrow iii). The necessary conditions for the novel DET are
satisfied in the red region in the $eV_{b}$-$V_{g}$ plain of Fig. 4(c). Note
that this complex excitation can only occur if $\tau _{spin}\gtrsim \Gamma
^{-1}$. Figure 4(a) shows the tunneling current spectrum of QD-2 between $N$
= 1 and 2 CB regions at $B$ = 0 T with no $V_{G}$-pulse. In addition to the
expected current step marked by arrow a associated with excitation to the ES$%
(2,1)$, we see an extra current step marked by arrow b associated with the
novel DET \cite{dcIincdec}. Note that the condition in dc measurement for
novel DET as well as the other tunneling processes is equivalent to that in
pulse measurements, by taking $\alpha V_{p}$ equivalent to $eV_{b}$.

In summary, we have discussed novel DET, which can lead to considerable
charge fluctuation, and ES-ES tunneling, which gives rise to significant
fluctuation in the total spin. These non-equilibrium transport processes
cannot be explained by orthodox Coulomb blockade theory, and arise from the
absence of effective spin relaxation inside a quantum dot. With
miniaturization of semiconductor devices selection rules for tunneling
transitions have to be considered when manipulating spin and charge in a
quantum dot.


\begin{thebibliography}{99}
\bibitem{higherVp} If the pulse height is made larger so that $U(1,\frac{1}{2%
})>$ $U(2,1)$ at $V_{G}=V_{h}$, the tunneling transitions from $(1,\frac{1}{2%
})$ to either $(2,0)$ or $(2,1)$ occurs more or less equivalently, but this
provides no net transient current.
\bibitem{Bdependence} The peak position reflects the orbital characteristic
of the corresponding two states, $(2,1)$ and $(3,\frac{1}{2})$ for peak b,
and $(2,0)$ and $(3,\frac{1}{2})$ for the GS-GS tunneling peak appearing
between $N=$ 2 and 3. We confirmed that the spacing between these two peaks
corresponds to $\Delta _{2}$ over the magnetic field range investigated (0
to 4 T).
\bibitem{dcIincdec} The current increases or decreases depending on the
tunneling conditions. This is partly explained in Ref. \cite{FujisawaPRB}.
\end{thebibliography}
\end{document}